\documentclass[12pt]{JHEP3}
\pdfoutput=1

\usepackage{graphicx,subfigure}
\usepackage{amsmath}

\newcommand{\bea}{\begin{eqnarray}}
\newcommand{\eea}{\end{eqnarray}}

\renewcommand{\k}{\kappa}

\newcommand{\m}{\mu}
\newcommand{\n}{\nu}
\renewcommand{\r}{\rho}
\newcommand{\s}{\sigma}

\newcommand{\pa}{\partial}

\newcommand{\rh}{r_H}

\title{\begin{center} 
On inhomogeneous holographic superconductors
\end{center}}
\author{
Massimo Siani
\\~
\\
Instituut voor Theoretische Fysica, Katholieke Universiteit Leuven,\\
Celestijnenlaan 200D B-3001 Leuven, Belgium.
\\~\\
\email{massimo.siani@fys.kuleuven.be}
}

\abstract{ We study a model describing a four-dimensional holographic superconductor whose properties depend non-trivially on a particular spatial direction, namely a Josephson junction. We analyze the parameter dependence of the condensate and compare it to the condensed matter expectations.
}

\preprint{}

\begin{document}

\section{Introduction}

Condensed matter systems often display a strongly coupled behavior near a quantum critical phase transition. The AdS/CFT correspondence \cite{Maldacena:1997re,Witten:1998qj,Gubser:1998bc} offers a powerful tool to study strongly coupled field theories. For reviews with particular emphasis on applications to condensed matter physics, see \cite{Hartnoll:2009sz,Herzog:2009xv,McGreevy:2009xe,Horowitz:2010nh}. The observation that a scalar field can break an abelian gauge symmetry near an AdS black hole horizon \cite{Gubser:2005ih,Gubser:2008px} has led to the recent construction of a model of holographic superconducting phase transition \cite{Hartnoll:2008vx}.

The simplest model of the holographic superconductor minimally couples an abelian vector field and a massive scalar field to the Einstein-Hilbert action with a negative cosmological constant. In the pioneering works \cite{Hartnoll:2008vx,Hartnoll:2008kx}, all the fields were supposed to depend on the radial AdS coordinate only, thus providing the gravity dual of a homogeneous superconductor. More recently, it has been shown that allowing the fields to non-trivially depend on the spatial coordinates of the boundary also gives rise to interesting phenomena in the dual field theory \cite{Horowitz:2011dz,Arean:2010xd}, namely it makes it possible to describe a $2+1$-dimensional holographic Josephson junction. The aim of the present paper is to extend their results to the $3+1$-dimensional case and to show that the condensed matter expectations also match in this case. This allows us to further extend the number of phenomena which can be described via holography.

The rest of the paper is organized as follows. In section \ref{gravitylag} we write the gravitational Lagrangian, derive the set of equations
of motion and review earlier results on the homogeneous solution to set the stage for the following discussion. Then, we move to the description of our inhomogeneous setting.
In section \ref{numerics} we present our numerical results for the superconducting phase and we compare them to the condensed matter expectations. We conclude in section \ref{conclusions}.

{\bf Note added:} immediately after our work was completed, we became aware of \cite{Wang:2011xx}, which has some overlap with our results.

\section{The gravity model} \label{gravitylag}
The minimal gravitational action describing a holographic model of a s-wave superconductor contains an abelian gauge field $A_\m$ and a charged minimally coupled scalar field $\phi$.
Allowing for a scalar mass term we then consider the action
\bea
S = \int d^5x \sqrt{-g} \left[ \frac{1}{2 \k^2} \left( R + \frac{12}{L^2} \right) - \frac{1}{4} F_{\m\n} F^{\m\n} - \left| D \phi \right|^2 - m^2 \left| \phi \right|^2 \right]
\label{eq:action}
\eea
where the value of the cosmological constant has been set to $12$, $\k^2=8\pi G_5$ sets the five-dimensional Newton constant,
\bea
F_{\m\n} &=& \pa_\m A_\n - \pa_\n A_\m \\
D_\m \phi &=& \nabla_\m \phi - i q A_\m \phi
\eea
are the gauge field strength and the covariant derivative respectively, and $\nabla_\m$ is the metric covariant derivative.
The gauge field has been rescaled so that its kinetic term does not involve any constant and $q$ is the charge of the scalar field in these units.

The Einstein, Maxwell and scalar equations of motion reads
\bea
\begin{split}
G_{\m\n} - \frac{6}{L^2} g_{\m\n} &= \k^2 T_{\m\n} \\
\nabla^\m F_{\m\n} &= q i \left( \phi^\ast ~ D_\m \phi - D_\m \phi^\ast ~ \phi \right) \\
\left( D^\m D_\m - m^2 \right) \phi &= 0
\end{split}
\label{eq:EOMs}
\eea
where $G_{\m\n}$ is the Einstein tensor and $T_{\m\n}$ is the energy-momentum tensor given by
\bea
T_{\m\n} = F_{\m\r} F_\n^{\ \r} - \frac{1}{4} g_{\m\n} \, F_{\r\s} F^{\r\s} + 2 D_{(\m} \phi \, D_{\n)} \phi^\ast - g_{\m\n} \left| D \phi \right|^2 - g_{\m\n} \, m^2 \left| \phi \right|^2
\eea

When the fields are constrained to only depend upon the radial coordinate, the model (\ref{eq:action}) describes a homogeneous holographic superconductor.
The common ansatz in that case is
\bea
\begin{split}
ds^2 &=  - f(r) e^{2 \n(r)} dt^2 + \frac{dr^2}{f(r)} + \frac{r^2}{L^2} d{\bf x}^2 \\
A_t &=h(r) \qquad A_i = 0 \quad i=r,x,y,z \\
\phi &= \phi(r)
\end{split}
\label{eq:homo}
\eea
where the scalar $\phi$ can be taken to be real without any loss of generality and the function $\n$ vanishes if the Newton constant $\k=0$. There exist
two phases separated by a critical temperature $T_c$. When $T > T_c$ the scalar field vanishes (normal phase) and the analytic solution is
\bea
\begin{split}
f(r) &= \frac{r^2}{L^2} \left(1- \frac{\rh^4}{r^4} \right) - \frac{2}{3} \m^2 \k^2 \frac{\rh^2}{r^2} \left( 1- \frac{\rh^2}{r^2} \right) \\
h(r) &= \m \left( 1- \frac{\rh^2}{r^2} \right) \qquad \qquad \n=0
\end{split}
\eea
Below $T_c$ the scalar field acquires a non-trivial profile and the corresponding solution describes the superconducting phase. Because we are left with a set of non-linear coupled differential equations, there is not a general analytic solution, but many numeric reliable methods are available. We will take (\ref{eq:homo}) to be one of our boundary conditions.

We are interested in finding an inhomogeneous solution of (\ref{eq:EOMs}), i.e. a solution in which the gauge and the scalar fields acquires a non-trivial dependence
on the spatial coordinates. To simplify our task, we consider the theory in the probe limit, namely $\k \to 0$; in that case the Einstein equations decouple from
the gauge-scalar sector and we can consistently consider the plane-symmetric ansatz for the metric
\bea
ds^2 = - f(r) dt^2 + \frac{dr^2}{f(r)} + \frac{r^2}{L^2} d{\bf x}^2
\eea
We impose the boundary condition $f(\rh)=0$ which determines, as a function of the horizon radius $\rh$ of the black hole, the explicit solution
\bea
f(r) = \frac{r^2}{L^2} \left(1-\frac{\rh^4}{r^4} \right)
\eea
and the Hawking temperature
\bea
T_H = \frac{\rh}{\pi L^2}
\eea
which is identified with the dual field theory temperature.

Without loss of generality we take the matter fields to depend on the spatial coordinate $x$ but not on $y$ and $z$. Thus, our ansatz reads
\bea
\begin{split}
A_i &= A_i(r,x) \quad i=t,r,x,y,z \qquad \qquad \phi &= \psi(r,x) e^{i \theta(r,x)}
\end{split}
\label{ansatzmatter}
\eea
In general, the phase $\theta$ of the scalar field cannot be set to zero, but it disappears from the equations of motion once
we shift the gauge field as $A_i \to m_i = A_i - \pa_i \theta$. Thus, a consistent subset of equations is given by only considering
a non-vanishing $m_t$ which satisfies
\bea
m_t^{\prime\prime} + \frac{3}{r} \, m_t^\prime + \frac{L^2}{r^2 f} \, \pa_x^2 m_t - \frac{2 \psi^2}{f} \, m_t = 0
\label{eq:mt}
\eea
where a prime indicates a derivative with respect to the radial coordinate $r$. Furthermore, the scalar equation reduces to
\bea
\psi^{\prime\prime} + \left( \frac{3}{r} + \frac{f'}{f} \right) \psi^\prime + \frac{L^2}{r^2 f} \, \pa_x^2 \psi + \left( -\frac{m^2}{f} + \frac{m_t^2}{f^2} \right) \psi = 0
\label{eq:psi}
\eea
The solution to the system (\ref{eq:mt})-(\ref{eq:psi}) is uniquely determined once we impose some boundary condition on the fields. We require that
at large values for $x$ the system reduces to the homogeneous holographic superconductor described by (\ref{eq:homo}). In that case, regularity imposes the gauge field $m_t$
to vanish at the horizon, while the scalar field has to satisfy $\phi(\rh)=(f^\prime(\rh)/m^2)\, \phi^\prime(\rh)$. At finite $x$,
the general asymptotic solution near the boundary is
\bea
\begin{split}
m_t(r,x) &\stackrel{r\to\infty}{\longrightarrow} \m(x) - \frac{\r(x)}{r^2} + \ldots \\
\psi(r,x) &\stackrel{r\to\infty}{\longrightarrow} \frac{\psi^{(+)}(x)}{r^\Delta} + \frac{\psi^{(-)}(x)}{r^{4-\Delta}} + \ldots
\end{split}
\label{eq:asymp}
\eea
where $m^2=\Delta(\Delta-4)$. According to the AdS/CFT dictionary, $\Delta$ is interpreted as the scaling dimension of the dual field theory operator which is acquiring a non-vanishing vacuum expectation value; thus,
$\psi^{(+)}$ is interpreted as the condensate value $\langle {\cal O}_\Delta \rangle$ of that operator and $\psi^{(-)}$ as its source. A spontaneous symmetry
breaking on the boundary theory requires that we choose $\psi^{(-)}=0$. The field theory chemical potential and charge density are provided
by $\m(x)$ and $\r(x)$ respectively. They are not independent parameters: once we set one of the two, the other quantity is determined by the solution to the system (\ref{eq:mt})-(\ref{eq:psi}). Thus, we set our last boundary condition by
choosing the following function for the chemical potential \cite{Horowitz:2011dz}
\bea
\m(x) = \m_\infty \left[ 1-\frac{1-\epsilon }{2 \text{Tanh}\left[\frac{l}{2 \sigma }\right]}\left(\text{Tanh}\left[\frac{x+\frac{l}{2}}{\sigma }\right]-\text{Tanh}\left[\frac{x-\frac{l}{2}}{\sigma }\right]\right) \right]
\label{eq:chempot}
\eea
The peculiar feature of (\ref{eq:chempot}) is the following. For $|x| \gtrsim l$ it approaches the constant value $\m_\infty$, while it quickly decreases at $x \simeq l$ to a constant value set by $\epsilon$. Thus, our model will describe two superconductors placed at large $x$ and separated by an insulating barrier of size $2l$\footnote{for low enough temperatures, the insulating barrier will switch to a superconducting one. We will work in a range of temperatures where this does not happen.}.
Although we do not expect our results to strongly depend on it, the function (\ref{eq:chempot}) is a well suited chemical potential to model an interesting class of condensed matter systems.

The normal phase of our inhomogeneous solution is defined by the solution in which the scalar field vanishes identically. In this case, the boundary
conditions we imposed only allow for the homogeneous solution
\bea
h(r) = \m \left( 1 - \frac{\rh^2}{r^2} \right)
\eea
where $\m$ is a constant.

The superconducting phase corresponds to the hairy black hole solution, in which the scalar field assumes a nontrivial profile. When the temperature is lower than a certain critical value, the AdS black hole solution discussed above produces an instability because the scalar effective mass violates the BF bound. Thus, at the critical value for the temperature a phase transition occurs.

In the rest of this paper, we use the following scaling symmetry of both the bulk Lagrangian and the equations of motion
\bea
&&r \to ar, t, x^i \to at, ax^i, L \to aL, q \to q/a, m \to m/a; \nonumber \\
&&r \to br, t,x^i \to t/b, x^i/b, f \to b^2f, m_t \to bm_t; \nonumber
\eea
to set $L=\rh=1$ in performing numerics. In our approximations, the value of $q$ does not affect our equations.

For concreteness, we choose $m^2 = -3$. In this way, the dimension of the boundary operator which condenses is set to $\Delta=3$\footnote{we could also choose
$\Delta=1$ in this case, but we do not expect our qualitative results to change with the dimension of the operator.}
and the value of the condensate is read from the asymptotic $r^{-3}$ falloff of the scalar field, when a nontrivial solution exists.

\section{The superconducting phase} \label{numerics}

\begin{figure}
\centering
 \includegraphics[width=.5\textwidth]{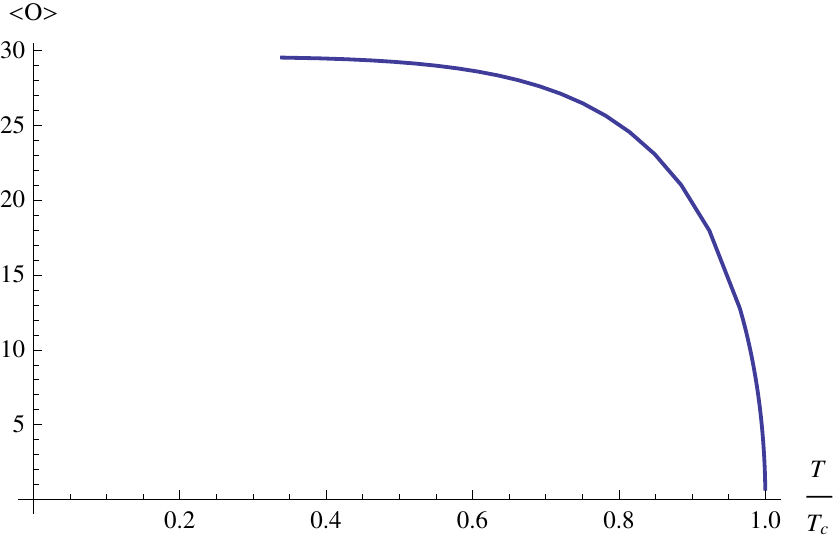}
\caption{The scalar condensate in the homogeneous solution as a function of the dimensionless temperature.}
\label{fig:homo}
\end{figure}

The solution at a finite value of the spatial coordinate $x$ strongly relies on the boundary solution at large $x$. The solution to the homogeneous solution
is found by expanding the fields and the equations (\ref{eq:mt}) and (\ref{eq:psi}) near the horizon $r=\rh$ and by dropping the spatial dependence. Thus,
we obtain
\bea
\begin{split}
m_t(r) &= m_t^\prime(\rh) \left[ (r-\rh) + \frac{(r-\rh)^2 \left(-6 + \psi(\rh)^2 \right)}{4 \rh} \right] + \ldots \\
\psi(r) &= \psi(\rh) \left[ 1 - \frac{3 (r-\rh)}{4 \rh} + \frac{(r-\rh)^2 \left(33 - m_t^\prime(\rh)^2 \right)}{64 \rh^2} \right] + \ldots
\end{split}
\eea
We assign an initial value to $m_t^\prime(\rh)$ and $\psi(\rh)$, numerically integrate the equations from the horizon to the boundary and use a shooting method
to seek a solution which satisfies $\psi^{(-)}=0$. Finally, we fit the asymptotic scalar solution to read off the operator condensate. The result is shown
in figure \ref{fig:homo}.

We solve the partial differential equations (\ref{eq:mt})-(\ref{eq:psi}) by compactifying the radial and spatial coordinate as $r^\prime = 1-1/r$ and $x^\prime = \tanh{(x/(4\sigma))}$ \cite{Horowitz:2011dz} and defining a grid on the resulting compact space; accordingly,
we substitute derivatives with their discrete versions. At $x=\pm \infty$ we impose that the fields approach the homogeneous solution discussed above, while at the
boundary of the AdS space we assign to $m_t$ the values of the chemical potential. In the
interior of the $(r,x)$ space we use the relaxation method. Once we have the value of the solution at each point of the grid, we interpolate them and fit the asymptotic functions
(\ref{eq:asymp}).

\begin{figure}
 \centering
\subfigure{\includegraphics[width=.48\textwidth]{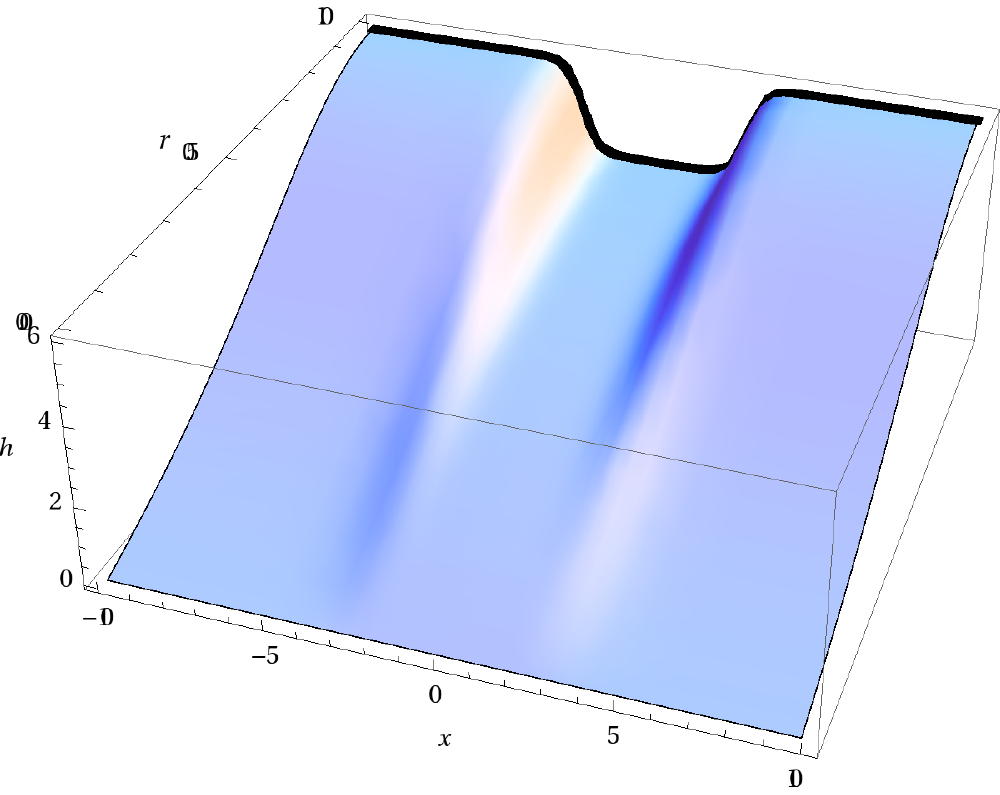}}
\subfigure{\includegraphics[width=.48\textwidth]{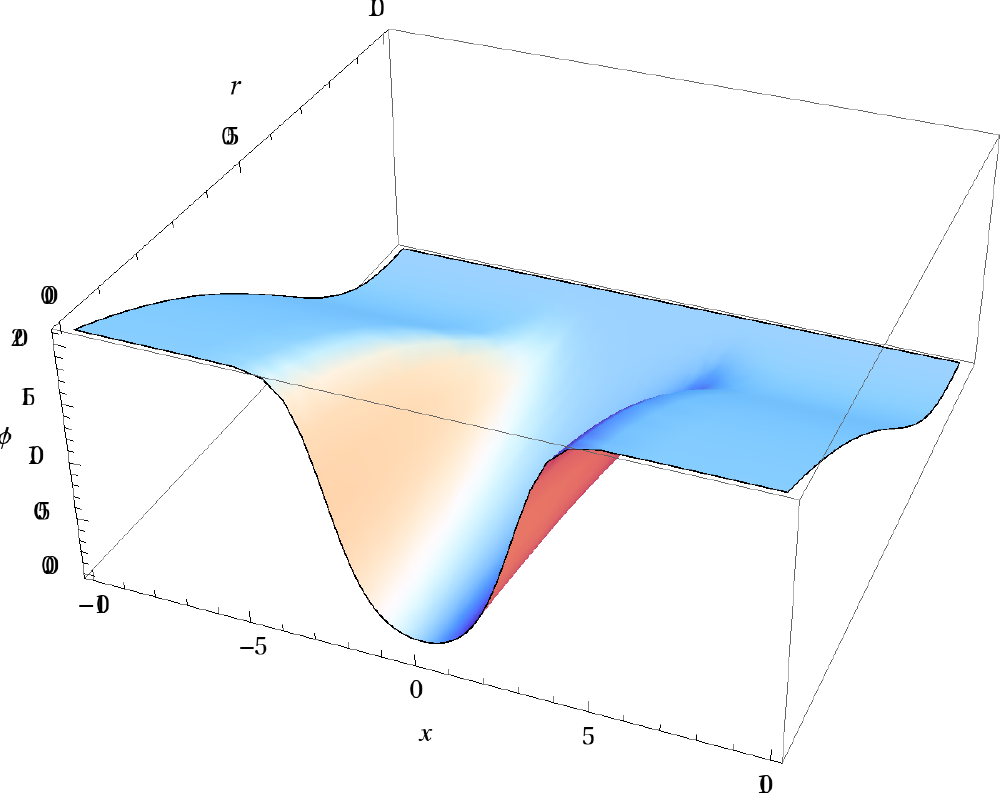}}
\caption{An example of the gauge (left) and scalar (right) field solution. The black line in the left plot corresponds to the chemical potential.}
\label{fig:inhomo}
\end{figure}

\begin{figure}
 \centering
 \subfigure{\includegraphics[width=.48\textwidth]{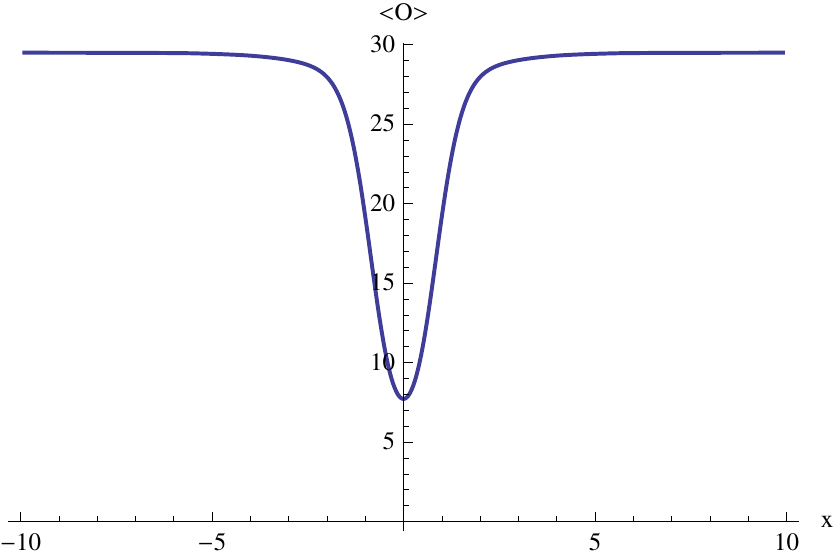}}
 \subfigure{\includegraphics[width=.48\textwidth]{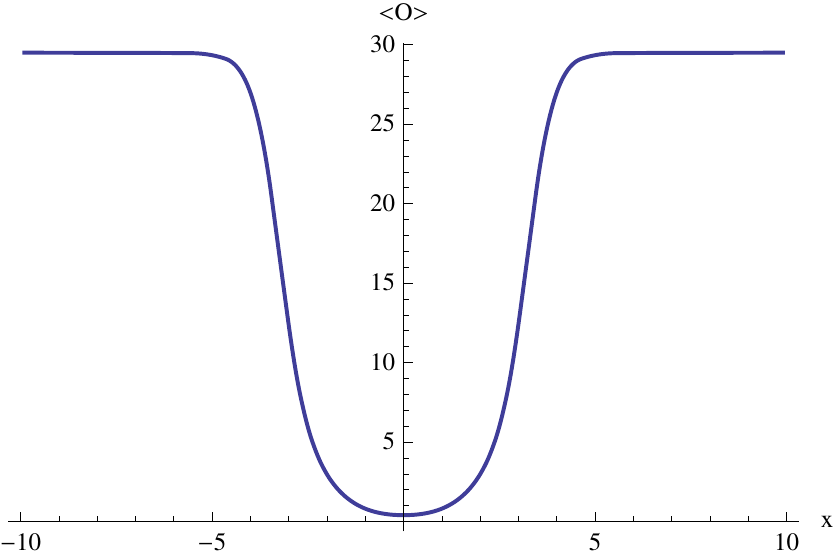}}
 \caption{The scalar condensate as a function of the spatial coordinate $x$ for $\epsilon=0.6$ and $l=1$ (left) and $l=6$ (right).}
 \label{fig:Ox}
\end{figure}

An example of our solution is shown in figure \ref{fig:inhomo}. We draw a black line corresponding to the boundary chemical potential and plot the solution between
$x=-10$ and $x=10$, where it is indistinguishable from the homogeneous solution. In fact, it can be seen that immediately $x$ is greater than $l$, the solution
becomes very flat along that direction. That is a consequence of the shape of the chemical potential we have chosen as our boundary condition.
A completely analogous argument also holds in the case of the scalar field. However, its shape distinguishes whether a non-trivial condensate appears in the dual
field theory. For a sufficiently large $l$ and low $\epsilon$ in (\ref{eq:chempot}) its value at $x=0$ stays very close to zero \footnote{the scalar field solution shown in figure \ref{fig:inhomo} can be well approximated as $\phi(r,x) \approx 4 \sigma {\rm Arctanh}^4 (x) \, \varphi(r)$, where $\varphi(r)$ is the homogeneous solution. We did not find any similar interpolating function either for the gauge field or for the scalar field with $l \lesssim 3$.}, resulting in a very low value for the condensate
in that region. We show this phenomenon in figure \ref{fig:Ox}. At $x\simeq l$ the condensate starts to dip below the homogeneous limit and it quickly decreases reaching
its minimum value at $x=0$. Figure \ref{fig:Ox} shows two different regimes, namely $l=1$ and $l=6$. In the former case the scalar condensate at $x=0$ is higher. Of course,
we expected such a behavior.

\begin{figure}[t]
 \centering
 \includegraphics[width=.5\textwidth]{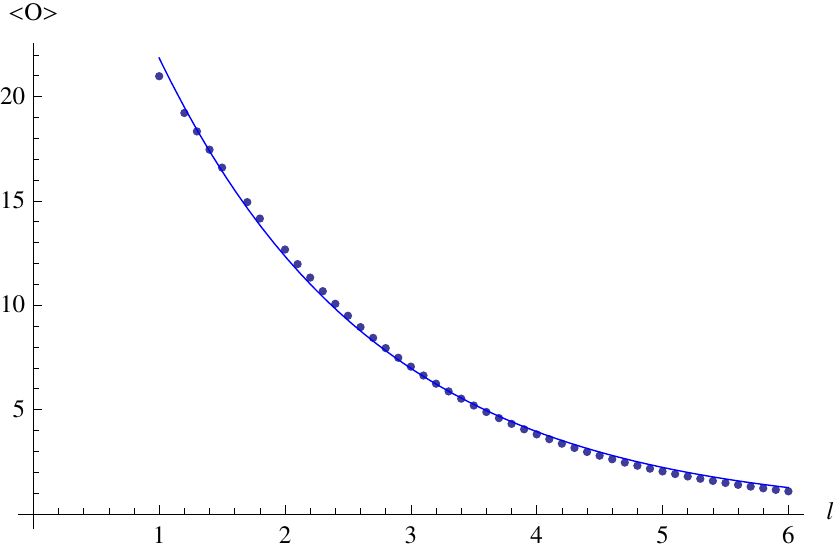}
 \caption{The scalar condensate as a function of the spatial coordinate $x$ for $l=6$ and $\epsilon=0.6$.}
 \label{fig:Ox0}
\end{figure}

We can go a bit further and find a function which nicely interpolates our numerical results at all $l$s. In the condensed matter literature, it is expected
that the scalar condensate at $x=0$ and the length $l$ are related by the simple relation
\bea
\frac{\langle {\cal O}_3 \rangle}{T_c^3} \Bigg|_{x=0} = A \, e^{-l/(2 \xi)}
\label{eq:Ox0}
\eea
where $\xi$ is the coherence length. Thus, we varied $l$ from $1$ to $6$ using an interval of $\Delta l=1/10$ to test the relation (\ref{eq:Ox0}) over a wide
range of parameters. Figure \ref{fig:Ox0} shows the results of our scan. The points represent our numerical results, while the interpolating line represents the
function (\ref{eq:Ox0}) where the parameters assume the values $A\simeq 38.6$ and $\xi \simeq 0.88$. Thus, we found a good agreement with the expected results.

\section{Conclusions} \label{conclusions}

We constructed a model of a four-dimensional inhomogeneous holographic superconductor in the probe limit. Our Lagrangian was the same as used in the gravity dual
of the homogeneous holographic superconductor, showing that the same model can holographically describe a large class of condensed matter systems, both in $2+1$
and in $3+1$ dimensions, by simply choosing appropriate boundary conditions for the fields in the ansatz. In the case of the boundary conditions we imposed, any anisotropies disappear at
large values of the spatial coordinate the fields depend on.
The lower dimensional counterpart was already studied in \cite{Horowitz:2011dz}, and although our results are not directly comparable to theirs, we
found many qualitative similarities.

The critical temperature and the scalar condensate are not affected by the anisotropies we considered, because we took the solution to approach the homogeneous
one at large distances. Nevertheless, we have shown that an appropriate choice of the boundary conditions contains many parameters corresponding to the
particular class of dual condensed matter systems we would like to model with our gravitational action. The full solution strongly depends on those parameters,
allowing for the finding of some relation among them and the dual field theory data that the gauge/gravity duality predicts. Those relations have been shown
to qualitatively agree with those expected from the condensed matter literature. This is, of course, a striking achievement.

We did not attempt to study the conductivity and the way inhomogeneities manifest themselves there. As long as the the electric field is treated as a small
external perturbation, it does not affect the superconducting background we constructed, but we expect it to depend on the spatial coordinates, even in the
probe limit, and reduce to the conductivity of the homogeneous solution at large $x$. In particular, it would be interesting to look for the fate of the frequency
gap over critical temperature ratio $\omega_g/T_c$ \cite{Horowitz:2008bn} in the interior of the $x$ space and to give the electric gauge field a non-trivial dependence on another transverse
spatial coordinate $z$ and the wavevector $k_z$ to study the second order expansion of the Green function. Such coefficient contains the information relative to the
magnetic permeability, and it can be shown that in the model we used here it contributes to achieving a negative refractive index.

We only considered the simplest model in which a quadratic potential is given to the scalar field which represents the gravity
dual of the boundary operator which condenses. Moreover, we only considered an abelian gauge field and the probe limit. The latter corresponds to give the bulk scalar
field a large charge under the abelian gauge field. Both because going beyond these assumptions can display new interesting phenomena and because
it seems difficult to derive a string theoretical model which reduces to this simple model, we think that it would be interesting to extend the present analysis
to include more ingredients.

\section*{Acknowledgments}

The author wishes to thank the organizers of the "Problemi Attuali di Fisica Teorica 2011" meeting where part of this work was completed.

\end{document}